\documentclass[letterpaper, 10 pt, conference]{ieeeconf}  

\IEEEoverridecommandlockouts                              
\usepackage[dvipdfmx]{graphicx} 
\usepackage{amsmath} 
\usepackage{amssymb}  
\usepackage{unit}
\usepackage{color}

\newtheorem{problem}{Problem}

\newcommand{\figref}[1]{Fig.~\ref{#1}}
\newcommand{\tabref}[1]{Tab.~\ref{#1}}
\def\eqref#1{Eq.~(\ref{#1})}
\def\secref#1{Sec.~\ref{#1}}

\renewcommand{\baselinestretch}{0.988}

\title{\LARGE \bf
Model Life Extension for Continuous Process:\\
Non-Invasive Correction of Model-Plant Mismatch with Regularization
}

\author{Yohei Kono$^1$ and Minoru Koizumi$^2$
\thanks{${}^1${Yohei~Kono} is with DX Engineering Research Department, Center for Digital Services, Hitachi,~Ltd.; 292 Yoshida-cho, Totsuka-ku, Yokohama, Japan:
        {\tt\small yohei.kono.un@hitachi.com}}%
\thanks{${}^2$Minoru Koizumi was with Digital Architecture Research Department, Center for Technology Innovation, Hitachi,~Ltd.; 292 Yoshida-cho, Totsuka-ku, Yokohama, Japan:
        {\tt\small m-koizu@krc.biglobe.ne.jp}}%
}

\begin{document}

\maketitle
\thispagestyle{empty}
\pagestyle{empty}

\begin{abstract}

In continuous process plants controlled by model predictive control,
model-plant mismatch (MPM), due to the aging of processes, causes degradation of control performance.
We propose a concept called Model Life Extension (MLE) and its implementation to mitigate this degradation in a non-invasive manner. 
The purpose of MLE is to continually update (re-identify) process models by using routine operating data on the assumption that the timescale of aging is much larger than the interval of excitation of reference signals.
We implemented MLE by estimating MPM via $\mathcal{L}_1$ regularized regression and by finding an optimal regularization parameter via cross-validation and showed through numerical experiments that an optimal parameter can exist and be found by cross-validation for a pilot-scale distillation column.
We then constructed the updated model based on the found parameter to demonstrate the possibility of correcting static-gain mismatch and transport-delay mismatch without injecting excitation signals to process inputs.

\end{abstract}

\section{INTRODUCTION}
\label{sec:intro}

Model predictive control (MPC) plays a vital role in the advanced control of industrial processes.
MPC involves using a dynamic model of a process to be controlled and simultaneously adjusts all process inputs to control all process outputs while accounting for their interactions \cite{Forbes2015}.
Hence, MPC has been widely introduced in large multi-input multi-output processes, specifically refining, chemical, and petrochemical \cite{Qin2003}.
Most of these processes are required and designed to operate 24 hours a day, which are referred to as {\em continuous processes}.

The performance of MPC closely depends on the accuracy of the model used.
Generally, the dynamics of processes can vary from when the model was designed (or MPC commissioned) because of the {\em aging} of processes, e.g., changes in feedstock rates and quality, degradation of instrumentation and process equipment, and changes in operating conditions \cite{Forbes2015,Qin2003}.
The gap between a model of a process and its actual dynamics, known as model-plant mismatch (MPM), degrades control performance \cite{S.Badwe2010,Tufa2016,Kalafatis2017}.
Also, since the determination of some design parameters for MPC (e.g., prediction and control horizons) closely depends on transport delay for process inputs,
MPC controllers might have to be re-designed once transport delay changes.
Thus, re-identification of process models is essential in MPC maintenance.

The re-identification of continuous processes has been implemented by closed-loop identification.
Since reference signals are related to production rates or quality, thus kept constant for several months \cite{Morita2018},
the variation in process inputs and outputs is not necessary to guarantee the accuracy of re-identified models.
To increase the variation, closed-loop identification frameworks involve the so-called step test \cite{Forbes2015},
i.e., injecting excitation signals (multiple steps) at process inputs.
Since the step test disturbs closed-loop operation and forces plant operators to monitor disturbed processes during data collection, the amplitude (or duration) of excitation signals should be as small (or short) as possible to the extent that the process models can be accurately re-identified.

Several groups of researchers and engineers have made progress in reducing such disturbance while guaranteeing the accuracy of re-identified models.
From our point of view, their studies are classified into two approaches. The first approach is to detect the precise location at which MPM appears, namely, input and output pairs that lead to poor prediction.
There is a variety of methods on how to detect this location: correlation analysis \cite{Tufa2016,Badwe2008}, Bayesian inference \cite{Huang2008}, and statistical tests \cite{Shigi}.
This detection reduces the number of inputs at which excitation signals are injected to reduce the disturbance with the step test. The second approach involves designing the test signals.
Bombois et al. \cite{Bombois2006} proposed an identification paradigm that minimizes the identification cost (experiment time or performance degradation due to signal injection) with prior bounds on identified models.
Miskovic et al. \cite{Miskovic2008} investigated how the excitation of limited (not all) references affects the accuracy of closed-loop identification of multi-variable systems.
Kalafatis et al. \cite{Kalafatis2017} developed a software package that automatically generates excitation signals where the disturbance by their injection and accuracy of re-identified models are balanced in an optimization framework.

However, to the best of our knowledge, the use of {\em routine operating data}---closed-loop operating data without any experiment---has not been investigated in the re-identification of continuous processes.
Theoretically, this might result from common frameworks of closed-loop identification based on pre-designed input signals.
From a practical aspect, this is caused by the lack of knowledge of control algorithms.
Vendors of MPC packages often make their MPC algorithms undisclosed to protect their intellectual property.
Hence, once MPC packages are deployed, it is difficult to understand and reduce the effect of noise feedback through MPC.
This indicates that the re-identification still relies on the injection of test signals.

We propose a concept called Model Life Extension (MLE) for mitigating the degradation in MPC performance in a non-invasive manner.
MLE involves taking particular note of the significant difference in the timescale between routine operating data and process aging; (i) at least one reference signal for process outputs excites per several months \cite{Morita2018}, and (ii) the timescale of aging is much larger than it.
On the basis of this difference, MLE is used to continually estimate MPM as precisely as possible using the data in which at least two different reference signals excite.
Although a model might only be partially re-identified by this, its accuracy might be improved.
By reflecting MPM in an original model, it is possible to extend the duration for which the accuracy is kept higher than its threshold for re-identification and thus reduce the number of step tests.

We constructed the concept of MLE and implemented it.
We formulate continuous processes as a linear, stable, and slowly varying system controlled by an MPC controller.
We implemented MLE by correcting MPM via $\mathcal{L}_1$ regularized regression \cite{Hastie} to avoid deterioration in corrected models' accuracy due to overfitting and fed back measurement noise.
The effectiveness of MLE closely depends on the appropriate choice of the regularization parameter.
Through numerical experiments, we determined the existence of an optimal parameter that
suppresses the regression error for a pilot-scale distillation column with two-types of MPM---static-gain mismatch and transport-delay mismatch---and found it by cross-validation \cite{Hastie}.
Through the above experiments, we show that parameter estimation without regularization might not accurately correct MPM because of the overfitting and fed back measurement noise and demonstrate that MLE enables the correction of MPM without injecting signals to process inputs.

The contributions of this note are as follows. The first is the development of a practical methodology of implementing MLE.
Estimating MPM and determining its hyperparameter does not require control laws.
Thus, MLE can be implemented in continuous processes where MPC has already been deployed.
The second contribution is to provide a practical means and numerical evidence for closed-loop identification with regularization.
There is a variety of theories and methods for system identification with regularization; see, e.g., \cite{Sjoberg1993,VanGestel2001} for parametric cases and \cite{Pillonetto2010,Chen2012,Chen2014,Formentin2018,Chiuso2020,Formentin2021} for non-parametric cases.
Our methodology based on the excitation of reference signals enables us to determine the regularization parameter in closed-loop systems.
To the best of our knowledge, we are the first show that an optimal regularization parameter for closed-loop identification can be found by the cross-validation.

The rest of this note is structured as follows.
In \secref{sec:system}, we discuss the formulating of continuous processes and the problem of MLE.
In \secref{sec:method}, we discuss the implementation of MLE via $\mathcal{L}_1$ regularized regression.
In \secref{sec:experiment}, we present our numerical experiments to investigate the existence of an optimal regularization parameter and the effectiveness of our implementation of MLE.
In \secref{sec:discussion}, we discuss the applicability of MLE and delineate its advantage and cost in comparison with conventional re-identification frameworks.
We conclude the note with prospects in \secref{sec:conclusion}.

\section{PROBLEM FORMULATION}
\label{sec:system}

We start with formulating the continuous processes addressed in this note.
Let process inputs be $\vct{u}(t)\in\mathbb{R}^m$, outputs be $\vct{y}(t)\in\mathbb{R}^p$, and measurement noise be $\vct{v}(t)\in\mathbb{R}^p$ at continuous time $t$.
Since it is often the case that continuous processes are stable and operated around their equilibrium, we consider the process represented by the following stable, linear, and time-varying system:
\begin{subequations}\label{eq:plant_ss}
\begin{align}
\dot{\vct{x}}(t) &= {\sf A}(t)\vct{x}(t) + \sum_{i=1}^{n\sub{d}}{\sf B}_i(t)\vct{u}(t-L_i)\\
\vct{y}(t) &= {\sf C}(t)\vct{x}(t) + \vct{v}(t)
\end{align}
\end{subequations}
where $\vct{x}(t)$ denotes the internal states of the process and $\dot{\vct{x}}(t)$ their derivative, and $L_i$ denotes the transport delay of mass or energy.
Since the delay depends on fluid-mixing efficiency and the length of multiple pipes, $L_i$ naturally takes multiple ($n\sub{d}$) values. Also, ${\sf A}(t)\in\mathbb{R}^{n \times n}$, ${\sf B}_i(t)\in\mathbb{R}^{n \times m}$, and ${\sf C}(t)\in\mathbb{R}^{p \times n}$ are system matrices where the temporal change of parameters represents the aging of the process.
We assume that the settling time $T\sub{s}$ for a step change in each input is of dozens of hours \cite{Tsuda2008,Ogawa2017}.
We also assume that the parameter change is characterized by a specific timescale $T_\mathrm{a}$, which is on the order of several years, i.e., $T\sub{a} \gg T\sub{s}$.
Therefore, the parameter change does not affect the dynamic behaviour of \eqref{eq:plant_ss} in its feedback control.
We approximately describe \eqref{eq:plant_ss} as a time-invariant system parametrized by $t$:
\begin{align}
{\sf G}(s; t) = {\sf C}(t)(s{\sf I}-{\sf A}(t))^{-1}\sum_{i=1}^{n\sub{d}}{\sf B}_i(t)\ee^{-L_i s},
\label{eq:plant_transfer_matrix}
\end{align}
where $s$ is the complex frequency and ${\sf I}_n$ the identity matrix of size $n$.
We regard ${\sf G}_0(s):={\sf G}(s;0)$ as {\em initial} dynamics and ${\sf G}(s;t) - {\sf G}_0(s)$ as MPM at time $t$.

Next, we consider a discrete-time model structure for MLE.
Most commercial MPCs are based on input-output models well fit with ${\sf G}_0(s)$, such as finite impulse response, finite step response, or (discrete-time) transfer function models \cite{Qin2003}.
To handle the change in transport delay, we use the following ARX model structure without pre-designed input delay:
\begin{align}
\vct{y}_{k+1} = {\sf R}
\left[
\begin{array}{c}
 \vct{y}_k\\
 \vct{u}_k\\
 \vdots\\
 \vct{y}_{k-d+1}\\
 \vct{u}_{k-d+1}
\end{array}
\right],
\label{eq:model_structure}
\end{align}
where $\vct{y}_k = \vct{y}(k\Delta t)$ and $\vct{u}_k = \vct{u}(k\Delta t)$ are sampled input and output with the period $\Delta t$ and ${\sf R} \in \mathbb{R}^{p\times d(m+p)}$ is the regression coefficient matrix, and $d$ is the model order common in inputs and outputs.
We set $d$ at a significant value (i.e., $d\Delta t$ is much larger than time-constants and transport delays) to reflect various types of MPM including transport delay mismatch in \eqref{eq:model_structure}.
Below, we explain how to derive \eqref{eq:model_structure} from other structures of process models.

We introduce the MPC controller for the process (\ref{eq:plant_ss}) based on \cite{Maciejowski2002}.
To clearly investigate the effects of updating process models, we ignore any restrictive condition for MPC formulation.
Let prediction and control horizons be $H\sub{p}$ and $H\sub{c}$, where $H\sub{p} > H\sub{c}$.
Since process dynamics are slow, the MPC controller manipulates $\vct{u}(t)$ every control period $\Delta t\sub{c} \geq \Delta t$.
Let $\vct{u}(l\Delta t\sub{c})$, $\vct{y}(l\Delta t\sub{c})$, $\vct{r}(l\Delta t\sub{c})$ be $\bar{\vct{u}}_l$, $\bar{\vct{y}}_l$, $\bar{\vct{r}}_l$, respectively.
For every discrete-time $l$, MPC predicts $\bar{\vct{y}}_{l + 1}, \cdots, \bar{\vct{y}}_{l + H\sub{p}}$ by integrating its process model (e.g., \eqref{eq:model_structure}).
The input rates $\Delta\vct{u}_1, \ldots, \Delta\vct{u}_{H\sub{c}}$ are used to define the input sequence
\begin{align}
  \bar{\vct{u}}_{l + i} = \begin{cases}
    \bar{\vct{u}}_{l + i - 1} + \Delta\vct{u}_i & (i = 1,\ldots, H\sub{c})\\
    \bar{\vct{u}}_{l + H\sub{c}} & (i = H\sub{c}+1,\ldots, H\sub{p})
\end{cases},
\label{eq:input_rate}
\end{align}
and $\{\vct{u}_k\}$ for integrating the process model is derived from the zero-order hold of the sequence.
Below, the predicted $\bar{\vct{y}}_{l + i}$ is denoted as $\bar{\vct{y}}_{l | i}$.
On the basis of the prediction, the MPC controller searches the optimal sequence $\{\Delta\vct{u}_1, \ldots, \Delta\vct{u}_{H\sub{c}}\}$ by minimizing the following objective function
\begin{align}
J = &\sum_{i=1}^{H\sub{p}}
(\bar{\vct{r}}_{l + i} - \bar{\vct{y}}_{l | i})^{\sf T}
{\sf Q_y} (\bar{\vct{r}}_{l + i} - \bar{\vct{y}}_{l | i}) + \sum_{i=1}^{H\sub{c}} \Delta\vct{u}_i^{\sf T} {\sf Q_u} \Delta\vct{u}_i,
\label{eq:control_objective}
\end{align}
where ${\sf Q_y}\in\mathbb{R}^{p \times p}$ and ${\sf Q_u}\in\mathbb{R}^{m \times m}$ denote positive semi-definite weight matrices.
Then, the MPC controller inputs the first value $\bar{\vct{u}}_{l + 1}$ to the target process ${\sf G}(s)$ at $t = (l+1)\Delta t\sub{c}$.
Here, at least one of the reference signals $\vct{r}(t)\in \mathbb{R}^p$ is excited not by an experiment (excitation test) but by a routine operation per several months because of the change of operating conditions \cite{Morita2018}.

For the above closed-loop system, we construct the problem statement of MLE.
Since the MPC controller uses an {\em initial} model well fit with ${\sf G}_0$, the year-scale change of ${\sf A}(t)$, ${\sf B}(t)$, and ${\sf C}(t)$ induces control errors.
The errors might be reduced by estimating MPM ${\sf G}(s;t)-{\sf G}_0(s)$ and correcting the initial model with MPM.
Recalling that control laws in commercial MPC packages are often undisclosed, it is desirable to estimate MPM without any knowledge of \eqref{eq:control_objective}.
Thus, the problem of MLE is summarized as follows:
\begin{problem}[MLE]
Consider the process (\ref{eq:plant_ss}) with its settling time $T\sub{s}$ controlled by an MPC controller
where at least one reference signal in $\vct{r}(t)$ excites per a specific time $T\sub{r}$.
Assume that the coefficient matrix ${\sf R} \in \mathbb{R}^{p\times d(m+p)}$ in \eqref{eq:model_structure}
well fit with the initial model ${\sf G}_0(s)$
and routine operating data $\{\vct{u}(t),\vct{y}(t) \ |\  0 < t < T\}$ where $T \gg T\sub{s}$ are given.
Then, derive the updated coefficient matrix $\hat{{\sf R}} \in \mathbb{R}^{p\times d(m+p)}$ that corrects MPM ${\sf G}(s) - {\sf G}_0(s)$ and prevents the deterioration in its accuracy due to overfitting and measurement noise feedback.
\end{problem}

\section{IMPLEMENTING MODEL LIFE EXTENSION}
\label{sec:method}

This section discusses the methodology for implementing MLE via $\mathcal{L}_1$ regularized regression.

We first investigate how to derive \eqref{eq:model_structure} from other types of models.
Coefficient matrix ${\sf R}$ can be sparse because input signals affect the process dynamics through multiple transport delays.
Hence, we identify ${\sf R}$ with $\mathcal{L}_1$ regularization using numerical data derived from original process models.
The identification procedure is as follows.
\begin{enumerate}
  \item Input a random signal vector $\vct{u}(t)$ ($0 \leq t \leq N_0 \Delta t$) to the original process model to derive output series $\vct{y}(t)$.
  \item Construct the following matrices
  \begin{align}
    {\sf X}_0 = \left[
    \begin{array}{ccc}
     \vct{y}_{d-1} & \cdots & \vct{y}_{N_0 - 1}\\
     \vct{u}_{d-1} & \cdots & \vct{u}_{N_0 - 1}\\
     \vdots  & \ddots & \vdots\\
     \vct{y}_0 & \cdots & \vct{y}_{N_0 - d}\\
     \vct{u}_0 & \cdots & \vct{u}_{N_0 - d}
    \end{array}
    \right],
    {\sf Y}_0 =
    \left[
     \vct{y}_d\ \ldots\ \vct{y}_{N_0}
    \right].\nonumber
   \end{align}
  \item Set $\lambda_0$ at a positive value much smaller than 1 and identify ${\sf R}$ by minimizing the following objective function:
  \begin{align}
    f_0 ({\sf X}_0, {\sf Y}_0) = \frac{\left\|
      {\sf Y_0} - {\sf R}{\sf X_0}
      \right\|\sub{F}^2}{2(N_0 - d)}
    + \lambda_0 \|\sf R\|_{\mathcal{L}_1},
    \label{eq:convert}
  \end{align}
  where $\|\cdot\|\sub{F}$ denotes the Frobenius ($L_2$) norm of matrix, and $\|\cdot\|_{\mathcal{L}_1}$ the $\mathcal{L}_1$ norm.
\end{enumerate}

On the basis of ${\sf R}$, we formulate the estimation of MPM with the routine operating data.
Since the reference $\vct{r}(t)$ changes as a step signal, updated model $\hat{\sf R}$ can be overfitted to the slow dynamics of the actual process.
The measurement noise $\vct{v}(t)$ is also fed back to $\vct{u}(t)$ via the MPC controller, inducing high variance\footnote{The variance is in terms of the well-known bias-variance trade-off \cite{Hastie}.} of $\hat{\sf R}$.
To avoid these issues, we again use $\mathcal{L}_1$ regularization.
Let us consider inputs and outputs sampled at $t = k_1\Delta t, \ldots, k_N\Delta t$ ($\gg T\sub{s}$) and the matrices ${\sf X} \in \mathbb{R}^{d(m+p) \times N}$ and ${\sf Y}\in\mathbb{R}^{p \times N}$ denoted as\footnote{The time-instances $k_1, \ldots, k_N$ are not necessarily consecutive.}
\begin{align}
 {\sf X} = \left[
 \begin{array}{ccc}
  \vct{y}_{k_1-1} & \cdots & \vct{y}_{k_N - 1}\\
  \vct{u}_{k_1-1} & \cdots & \vct{u}_{k_N - 1}\\
  \vdots  & \ddots & \vdots\\
  \vct{y}_{k_1-d} & \cdots & \vct{y}_{k_N - d}\\
  \vct{u}_{k_1-d} & \cdots & \vct{u}_{k_N - d}
 \end{array}
 \right],
 {\sf Y} =
 \left[
  \vct{y}_{k_1}\ \ldots\ \vct{y}_{k_N}
 \right].\label{eq:data_matrix}
\end{align}
We then formulate MPM as the residual $\Delta {\sf R}$ derived by minimizing the following regularized loss:
\begin{align}
f_\lambda ({\sf X}, {\sf Y}, {\sf R}) = \frac{1}{2N}\left\|
{\sf Y} - ({\sf R}+\Delta {\sf R}){\sf X}
\right\|\sub{F}^2 + \lambda \|\Delta {\sf R}\|_{\mathcal{L}_1},
\label{eq:regression}
\end{align}
where $\lambda$ is the regularization parameter.
The second term is the penalty that prevents overfitting to the slow dynamics and decreases the MPM variance.
Once $\Delta {\sf R}$ is accurately estimated,
the updated matrix $\hat{{\sf R}}_\lambda := {\sf R} + \Delta {\sf R}$ is constructed and then implemented in the MPC controller.

We investigated how to choose $\lambda$.
As $\lambda$ takes a more significant value, the fit of coefficient matrix $\hat{{\sf R}}$ to output matrix ${\sf Y}$ becomes worse, known as the bias-variance trade-off \cite{Hastie}.
Then an optimal value of $\lambda$ that maximizes the fit of $\hat{\sf{R}}$ to unknown outputs can exist.
The existence of an optimal value has been implied in linear system identification; see, e.g., \cite{Sjoberg1993,Chen2012}.
For identifying finite impulse models, Chen et al. \cite{Chen2012} showed that the optimal regularization parameter is given by the impulse response of actual processes and variance of measurement noise\footnote{The variance is of signals.}.
They also presented various numerical results in which the optimal parameter is derived by two-fold cross-validation.
Inspired by the results, we assume that the optimal $\lambda$ exists for the MPM estimation (\ref{eq:regression}) and can be found by two-fold cross-validation.
Based on the method proposed in \cite{Chen2012}, we conduct two-fold cross-validation as follows.
\begin{enumerate}
\item Prepare two datasets $\mathcal{D}_1 =\{{\sf X}_1, {\sf Y}_1\}$
and $\mathcal{D}_2 = \{{\sf X}_2, {\sf Y}_2\}$.
\item Estimate coefficient matices $\hat{\sf R}_\lambda$ using $\mathcal{D}\sub{1}$ for different values of $\lambda$
and form the loss $f_\lambda ({\sf X}_2, {\sf Y}_2)$ for these matrices.
\item Exchange $\mathcal{D}_1$ and $\mathcal{D}_2$ and form the loss $f_\lambda ({\sf X}_1, {\sf Y}_1)$ in the same manner as above.
\item Pick the value of $\lambda$ that minimizes $f_\lambda ({\sf X}_1, {\sf Y}_1)+f_\lambda ({\sf X}_2, {\sf Y}_2)$.
\item Construct the whole dataset $\mathcal{D} = \{{\sf X}, {\sf Y}\}$ from $\mathcal{D}\sub{1}$ and $\mathcal{D}\sub{2}$
and re-estimate the matrix $\hat{\sf R}_\lambda$ by minimizing $f_\lambda ({\sf X}, {\sf Y}, {\sf R})$ with the picked $\lambda$.
\end{enumerate}

\begin{figure*}[!t]
  \begin{center}
   \includegraphics[width=0.7\hsize]{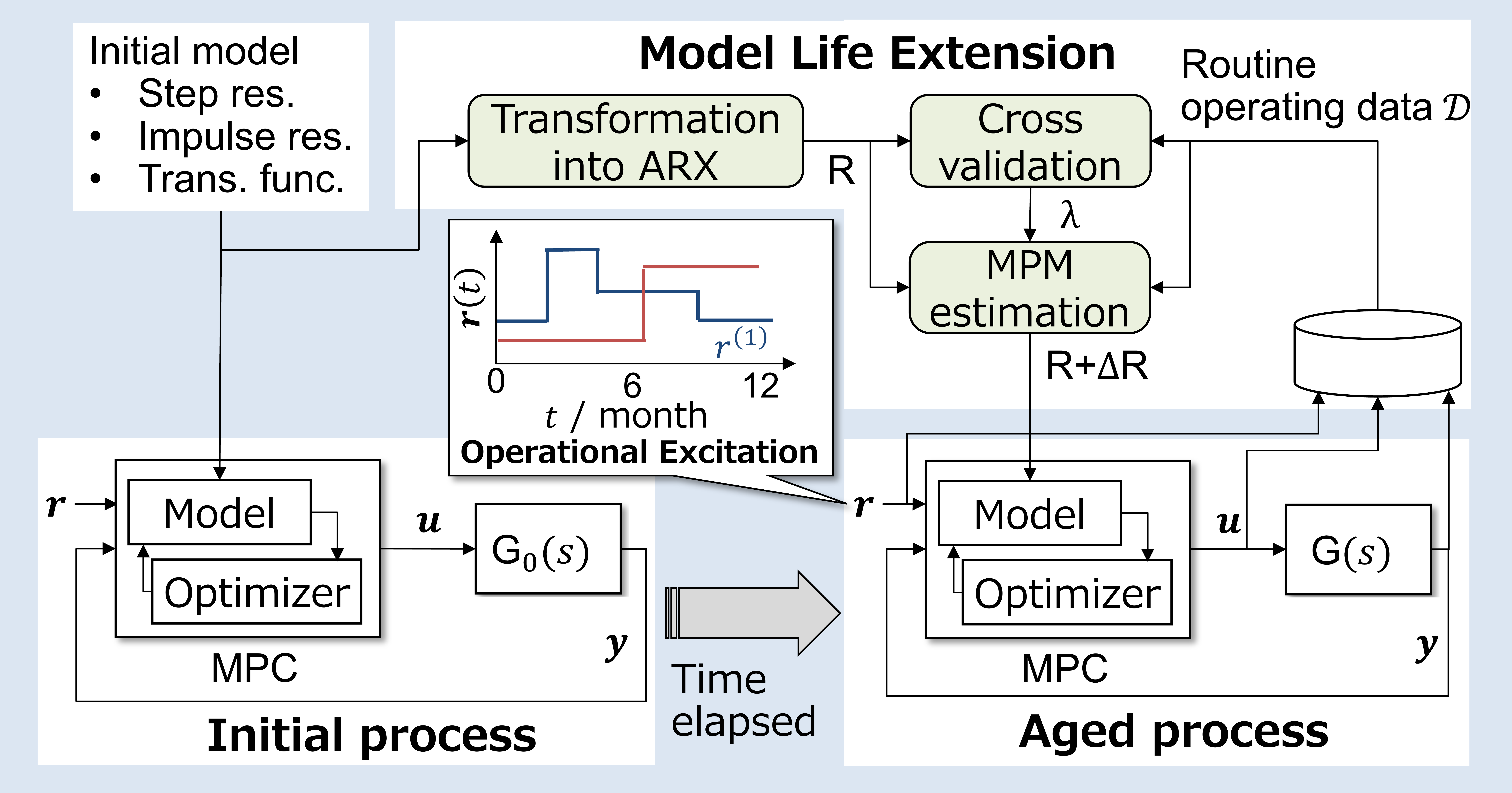}
   \caption{Conceptual diagram of MLE.}
   \label{fig:conceptual_diagram}	
  \end{center}
 \end{figure*}

The remaining problem lies in how to prepare the datasets $\mathcal{D}_1$ and $\mathcal{D}_2$.
Conceptually, it is desirable for model validation that prepared datasets reflect different dynamic properties of the process.
To this end, we utilize the excitation of reference signals.
As stated in \secref{sec:system},
while the aging timescale of processes is several years,
at least one reference signal excites per several months.
It naturally follows that the different reference signals can excite
while the process (\ref{eq:plant_ss}) is regarded as a time-invariant system.
Therefore, we propose to collect time-series data around the time each reference signal excites.
The $\mathcal{D}_1$ is constructed by a time series around when one(s) of the reference signals excites.
Likewise, $\mathcal{D}_2$ is constructed by a time series around when the other(s) excites.
We formulate a practical way of determining the length of the collected time-series based on a timescale of process dynamics in the next section.
The above MLE implementation is illustrated in \figref{fig:conceptual_diagram}.
%

\section{NUMERICAL EXPERIMENTS}
\label{sec:experiment}

We conducted numerical experiments to evaluate our implementation of MLE.
Note that all experiments were based on continuous-time models.
The derived data were then sampled with the fixed time-step $\Delta t = 0.2$\,min.

\subsection{CLOSED-LOOP SYSTEM CONFIGURATION}

\begin{figure}[!tb]
  \begin{center}
  \includegraphics[width=0.7\hsize]{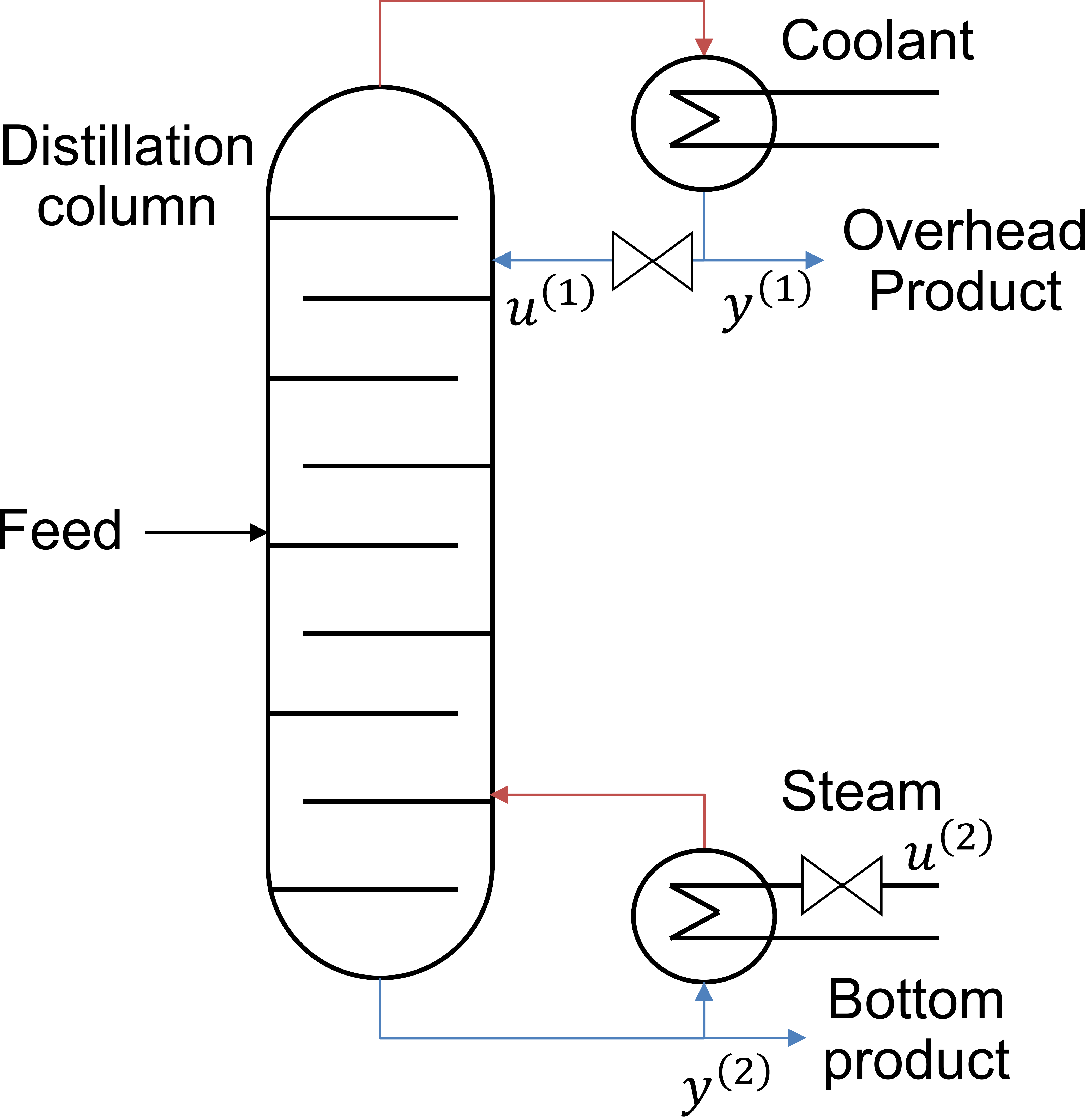}
  \caption{Pilot-scale distillation column.}
  \label{fig:distillation_column}	
  \end{center}
  \end{figure}

Let us introduce the target process and formulate its aging.
We address a distillation column, one of the essential processes for refineries and chemical plants.
Figure~\ref{fig:distillation_column} shows a schematic diagram of a pilot-scale binary distillation column \cite{wood1973}, where feed liquid is separated into two types of products by heating (evaporating) leaked liquid and cooling (condensing) vapour.
Here, $y^{(1)}$ denotes the concentration of the overhead product, $y^{(2)}$ the concentration of the bottom product, $u^{(1)}$ the flow rate of reflux from the condenser, and $u^{(2)}$ the flow rate of steam through the reboiler.
In the rest of this note, the feed flow rate was fixed at 2.45\,lb/min. Around the equilibrium condition \{$y^{(1)} = 96.0$\,wt\%, $y^{(2)} = 0.50$\,wt\%, $u^{(1)} = 1.95$\,lb/min, $u^{(2)} = 1.71$\,lb/min\}, the dynamic response of the concentration can be described as the transfer-function matrix \cite{wood1973}:
\begin{align}
  \left[
  \begin{array}{c}
    y^{(1)}\\
    y^{(2)}
  \end{array}
  \right] =
 \left[
 \begin{array}{cc}
  {\displaystyle\frac{K_{11}\ee^{-L_{11}s}}{1+T_{11}s}} & {\displaystyle\frac{K_{12}\ee^{-L_{12}s}}{1+T_{12}s}}\\
  {\displaystyle\frac{K_{21}\ee^{-L_{21}s}}{1+T_{21}s}} & {\displaystyle\frac{K_{22}\ee^{-L_{22}s}}{1+T_{22}s}}
 \end{array}
 \right]
 \left[
  \begin{array}{c}
    u^{(1)}\\
    u^{(2)}
  \end{array}
  \right],
 \label{eq:column_model}
\end{align}
where the parameters are summarized in \tabref{tab:model_parameter}.
We will use the matrix in \eqref{eq:column_model} as the initial dynamics ${\sf G}_0(s)$.

To verify the effectiveness of MLE, we considered two types of MPM of the process (\ref{eq:column_model}).
The first MPM is a static gain mismatch, which has the most impact on the degradation in MPC performance \cite{Tufa2016}.
We assume that the pipe between the column and condenser has deteriorated so that the static gain mismatch appears as follows:
\begin{align}
{\sf G}\sub{gain}(s) =
\left[
\begin{array}{cc}
 {\displaystyle\frac{(K_{11} + \Delta K_{11})\ee^{-L_{11}s}}{1+T_{11}s}} & {\displaystyle\frac{K_{12}\ee^{-L_{12}s}}{1+T_{12}s}}\\
 {\displaystyle\frac{(K_{21} + \Delta K_{21})\ee^{-L_{21}s}}{1+T_{21}s}} & {\displaystyle\frac{K_{22}\ee^{-L_{22}s}}{1+T_{22}s}}
\end{array}
\right].
\label{eq:column_model_aged}
\end{align}
As the second MPM, we address a transport delay mismatch, which might not be handled with the model structure that requires pre-designed input delay.
We modified $L_{21}$ and $L_{22}$ as follows:
\begin{align}
  {\sf G}\sub{delay}(s) = 
  \left[
  \begin{array}{cc}
    {\displaystyle\frac{K_{11}\ee^{-L_{11}s}}{1+T_{11}s}} & {\displaystyle\frac{K_{12}\ee^{-(L_{12} + \Delta L_{12})s}}{1+T_{12}s}}\\
    {\displaystyle\frac{K_{21}\ee^{-L_{21}s}}{1+T_{21}s}} & {\displaystyle\frac{K_{22}\ee^{-(L_{22} + \Delta L_{22})s}}{1+T_{22}s}}
  \end{array}
  \right].
  \label{eq:column_model_delayed}
  \end{align}
We made these mismatches more significant than their thresholds proposed in \cite{Tufa2016},
above which the degradation in MPC performance is not acceptable.
The parameter change due to these mismatches is summarized in \tabref{tab:model_parameter}.

We delineate the method for closed-loop simulations to generate datasets on $\vct{u}$ and $\vct{y}$.
The MPC controller was implemented using MATLAB\footnote{MATLAB is a registered trademark of MathWorks, inc.} MPC toolbox.
The controller uses ${\sf G}_0(s)$ as its process model.
The weight matrix ${\sf Q_y}\in\mathbb{R}^p$ was set at $\mathrm{diag(0.2, 0.2)}$
and ${\sf Q_u}\in\mathbb{R}^m$ at $\mathrm{diag(0.1, 0.1)}$\footnote{$\mathrm{diag(\cdot)}$ denotes a diagnal matrix.}.
The control period $\Delta t\sub{c}$ was set at 1\,min,
prediction horizon $H\sub{p}$ at 30, larger than $L_{12} + T_{12}$,
and control horizon $H\sub{c}$ at 5.
With this MPC controller and the process dynamics ${\sf G}\sub{gain}$ (or ${\sf G}\sub{delay}$),
we conducted two closed-loop simulations for $t \in [0, 1000]$.
Let reference signals be $\vct{r} := [r^{(1)}\ r^{(2)}]^{\sf T}$
and $U(t)$ be the unit step function.
We set $(r^{(1)}(t), r^{(2)}(t)) = (U(t - t\sub{r}), 0)$ in the first simulation
and $(r^{(1)}(t), r^{(2)}(t)) = (1, U(t - t\sub{r}))$ in the second,
where $t\sub{r} = 500$.
The excitation of the reference signals represents the change in the manufacturing plan of two (overhead and bottom) products.
For both the two simulations, measurement noise $\vct{v}$ was fixed at a two-dimensional Gaussian random signal with a variance of 0.001.
\begin{table}[tb]
  \begin{center}
  	\caption{Model parameters and their changes.}
  	\label{tab:model_parameter}
  	\begin{tabular}{c|ccc|cc}
    	$(i,j)$ & $T_{ij}$ / min & $L_{ij}$ / min & $K_{ij}$ & $\Delta K_{ij}$ & $\Delta L_{ij}$ / min \\\hline
    	$(1,1)$ & 16.7 & 1.0 & 12.8 & -6.4 & ---\\
    	$(1,2)$ & 21.0 & 3.0 & -18.9 & --- & 4.0\\
    	$(2,1)$ & 10.9 & 7.0 & 6.6 & -3.3 & ---\\
    	$(2,2)$ & 14.4 & 3.0 & -19.4 & --- & 4.0\\\hline
  	\end{tabular}
  \end{center}
\end{table}

\subsection{EXPERIMENTAL METHOD}

Let us explain how to implement MLE for the closed-loop system.
We inputted a two-dimensinal Gaussian random signal with a variance of 1
and a length of $N_0 \Delta t = 10000$\,min to the initial model ${\sf G}_0(s)$
and fixed $\lambda_0$ at $1.0\times 10^{-6}$ in \eqref{eq:convert}
to identify ${\sf R}$.
The model order $d$ was set at 150, larger than $(T_{ij} + L_{ij}) / \Delta t$.
The minimizing problems (\ref{eq:convert}) and (\ref{eq:regression}) were solved using the MATLAB command \texttt{lasso}.
The duration for collecting $\mathcal{D}_1$ and $\mathcal{D}_2$ was selected to be much larger than $T_{ij} + L_{ij}$.
We constructed the dataset $\mathcal{D}_1$ (or $\mathcal{D}_2$) by extracting $\{\vct{u}(t), \vct{y}(t) | t \in [t\sub{r} - 200,\ t\sub{r} + 200]\}$ from the first (or second) simulation data.
Candidates of regularization parameter $\lambda$ were in $\Lambda = \{0, 10^{-6}, 2\times 10^{-6}, \ldots, 10^{-2}\}$,
from which the minimizer $\lambda^*$ of cross validation error was chosen. 

For both ${\sf G}\sub{gain}(s)$ and ${\sf G}\sub{delay}(s)$, we evaluated the effectiveness of MLE through the following two steps.
\begin{enumerate}
\item Show the possibility that an optimal $\lambda$ exists and the validity of $\lambda^*$ found with MLE compared with the optimal value. 
\item Investigate how MLE reduces the mismatch of static gain (or transport delay) with the found $\lambda^*$.
\end{enumerate}
In step (1), we derived multiple matrices $\hat{\sf R}_\lambda$ by minimizing \eqref{eq:regression} for different $\lambda \in \Lambda$
and quantified their deviation from the true dynamics.
Since the step response is one of the most common benchmarks of process models \cite{Forbes2015},
the deviation was quantified by the mean-squared error between the step response of the true dynamics $\mathcal{M} \in \{ {\sf G}\sub{gain}, {\sf G}\sub{delay}\}$ and that of its model $\hat{\mathcal{M}} \in \{{\sf G}_0(s), \hat{\sf R}_\lambda\}$:
\begin{align}
E(\mathcal{M}, \hat{\mathcal{M}}) = \sum_{i,j\in \{1,2\}}\sum_{k=0}^{N}
\left(
\phi_k^{ij} (\mathcal{M})  - \phi_k^{ij}(\hat{\mathcal{M}})
\right)^2,
\end{align}
where $\{\phi_0^{ij}(\cdot), \ldots, \phi_N^{ij}(\cdot)\}$ represents the discrete-time step response from $u^{(i)}$ to $y^{(j)}$ for the true dynamics $\mathcal{M}$ or its model $\hat{\mathcal{M}}$.
The duration $N\Delta t$ was set at 100\,min.
While $E(\mathcal{M}, {\sf G}_0(s))$ takes a scalar value,
we illustrated $E(\mathcal{M}, \hat{\sf R}_\lambda)$ as a function of $\lambda$ to show the existence of an optimal $\lambda$ and the validity of $\lambda^*$.
In step (2), we compared the dynamic response curves of multiple models.
We investigated how step response curves of $\hat{\sf R}_{\lambda^*}$ are improved from those of ${\sf G}_0$ and how $\hat{\sf R}_0$ deteriorates from ${\sf G}_0$.
The accuracy of corrected static gain was quantified via the heights of the step response curves.
We also showed impulse response curves of $\hat{\sf R}_{\lambda^*}$ to calculate the corrected transport delay.

\subsection{EXPERIMENTAL RESULTS}

\begin{figure}[!tb]
\begin{center}
\includegraphics[width=\hsize]{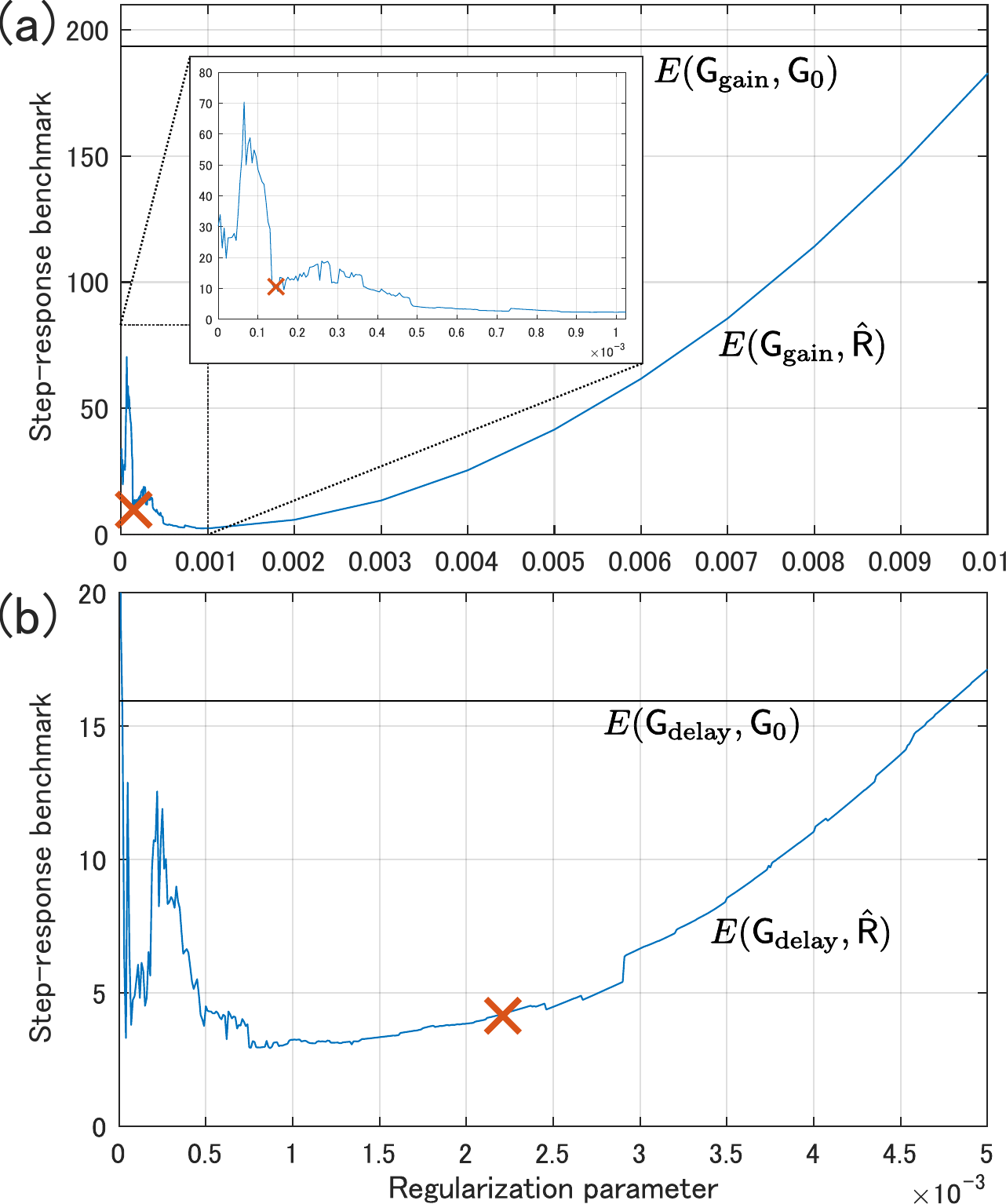}
\caption{Step-response benchmarks for (a) ${\sf G}\sub{gain}$ and (b) ${\sf G}\sub{delay}$.
Blue curves denote (a) $E({\sf G}\sub{gain}, \hat{\sf R}_\lambda)$ and (b) $E({\sf G}\sub{delay}, \hat{\sf R}_\lambda)$,
black lines (a) $E({\sf G}\sub{gain}, {\sf G}_0)$ and (b) $E({\sf G}\sub{delay}, {\sf G}_0)$, and
red cross regularization parameter $\lambda^*$ found by cross-validation.
In the panel~(a), the domain encompassed by the broken lines is enlarged in the top subfigure.}
\label{fig:regularized_loss}	
\end{center}
\end{figure}

Figure~\ref{fig:regularized_loss} shows the step-response benchmarks for (a) ${\sf G}\sub{gain}$ and (b) ${\sf G}\sub{delay}$.
The blue curves denote (a) $E({\sf G}\sub{gain}, \hat{\sf R}_\lambda)$ and (b) $E({\sf G}\sub{delay}, \hat{\sf R}_\lambda)$,
and the black lines (a) $E({\sf G}\sub{gain}, {\sf G}_0)$ and (b) $E({\sf G}\sub{delay}, {\sf G}_0)$.
For both panels (a) and (b), the red crosses correspond to $\lambda^*$ found by the cross-validation.
The $E({\sf G}\sub{gain}, \hat{\sf R}_\lambda)$ (or $E({\sf G}\sub{delay}, \hat{\sf R}_\lambda)$) took large values for small $\lambda$,
indicating overfitting.
Also, $E({\sf G}\sub{gain}, \hat{\sf R}_\lambda)$ (or $E({\sf G}\sub{delay}, \hat{\sf R}_\lambda)$) was almost equal to (or larger than) $E({\sf G}\sub{gain}, {\sf G}_0)$ (or $E({\sf G}\sub{delay}, {\sf G}_0)$),
which means that underfitting occurred.
Here, $E({\sf G}\sub{gain}, \hat{\sf R}_\lambda)$ and $E({\sf G}\sub{delay}, \hat{\sf R}_\lambda)$ could be minimized around $\lambda = 1.0\times 10^{-3}$,
indicating that an optimal regularization parameter exists for both mismatches.
The found parameters $\lambda^*$ are in the same order as $1.0\times 10^{-3}$, showing their validity.

Next, we illustrate how MLE reduces static-gain mismatch.
\begin{figure}[!tb]
\begin{center}
\includegraphics[width=\hsize]{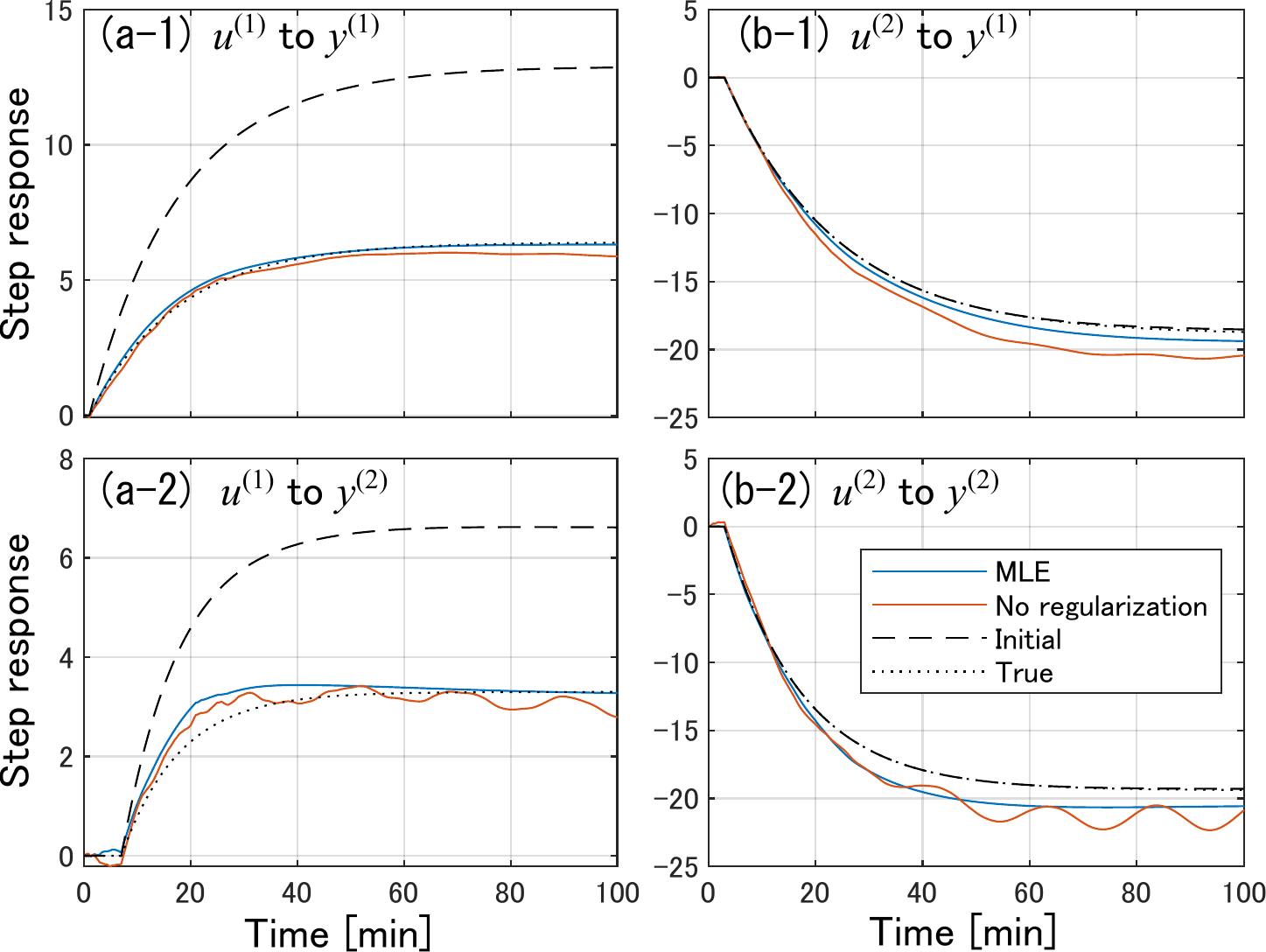}
\caption{Step-response curves derived in experiment for ${\sf G}\sub{gain}$.
Dashed lines represent response of initial dynamics ${\sf G}_0$, dotted lines true dynamics ${\sf G}\sub{gain}$, blue lines ${\sf R}_{\lambda^*}$ estimated using MLE, and red lines ${\sf R}_0$ estimated without regularization.}
\label{fig:step_response_static}	
\end{center}
\end{figure}
Figure~\ref{fig:step_response_static} shows step-response curves derived in the experiment for ${\sf G}\sub{gain}$.
The dashed lines represent the response of the initial dynamics ${\sf G}_0$, dotted lines the true dynamics ${\sf G}\sub{gain}$, blue lines ${\sf R}_{\lambda^*}$ estimated using MLE, and red lines ${\sf R}_0$ estimated without regularization.
First, undesired oscillation appeared in the step response of ${\sf R}_0$, while ${\sf R}_{\lambda^*}$ did not induce any oscillation.
Thus, parameter-estimation methods without regularization (e.g., online estimation) might deteriorate the accuracy of process models.
Also, in panels (a-1) and (a-2), the blue lines are close to the dotted line, showing that MLE enables accurate correction of the static-gain mismatch.
By comparing the final outputs, the static-gain mismatch between ${\sf R}_{\lambda^*}$ and ${\sf G}\sub{gain}$ was calculated at 9.33$\times 10^{-2}$ for $(i,j)=(1,1)$ and 5.15$\times 10^{-2}$ for $(i,j)=(2,1)$, sufficiently smaller than the original mismatch $|\Delta K_{11}| = 6.4$ and $|\Delta K_{21}| = 3.3$.

\begin{figure}[!tb]
  \begin{center}
  \includegraphics[width=\hsize]{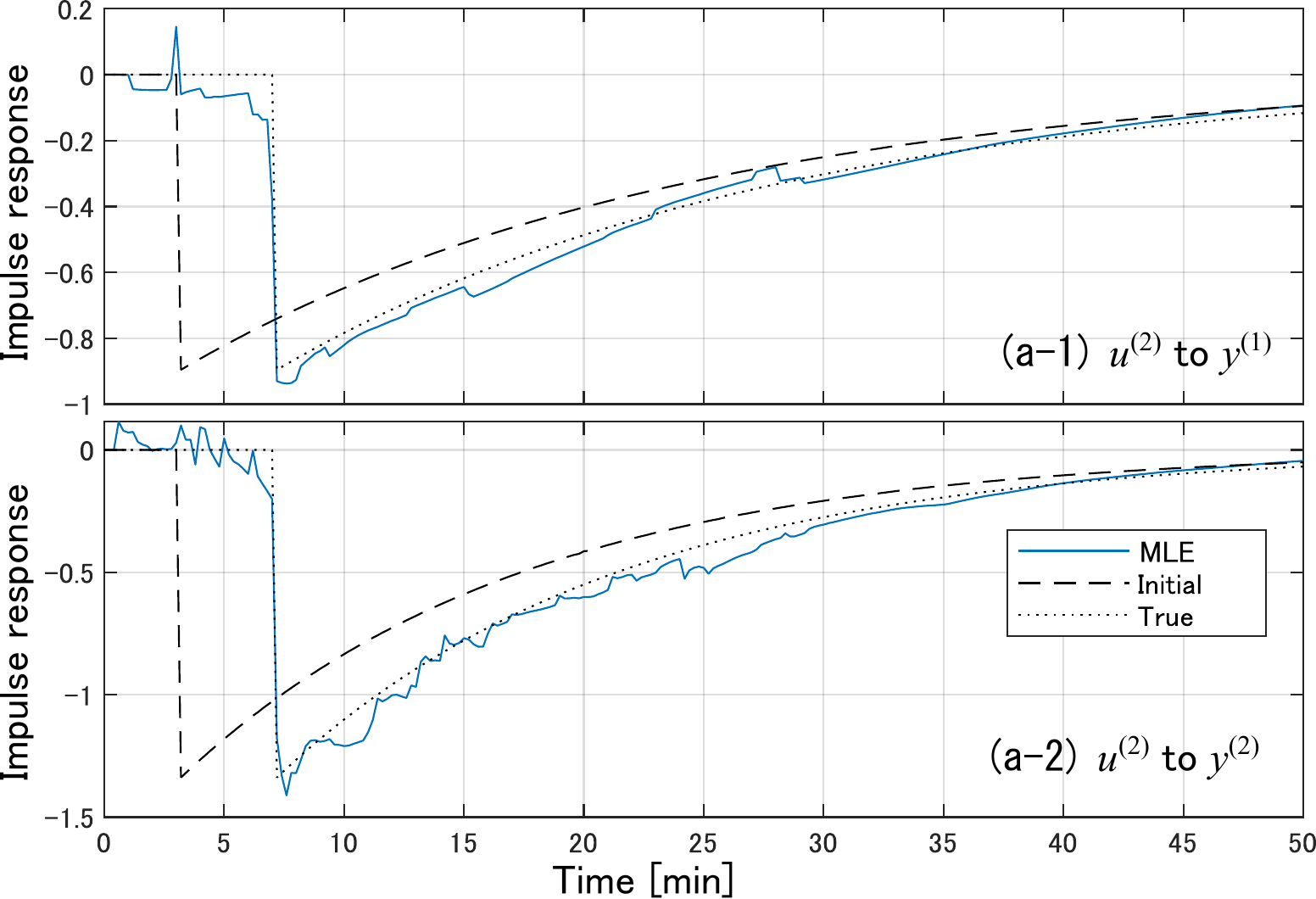}
  \caption{Impulse-response curves derived in experiment for ${\sf G}\sub{delay}$.
  Dashed line represents response of initial dynamics ${\sf G}_0$, dotted line true dynamics ${\sf G}\sub{delay}$, and blue line ${\sf R}_{\lambda^*}$ estimated by MLE.}
  \label{fig:Impulse_response_delay}	
  \end{center}
\end{figure}

We then investigate transport delay mismatch as above.
\begin{figure}[!tb]
  \begin{center}
  \includegraphics[width=\hsize]{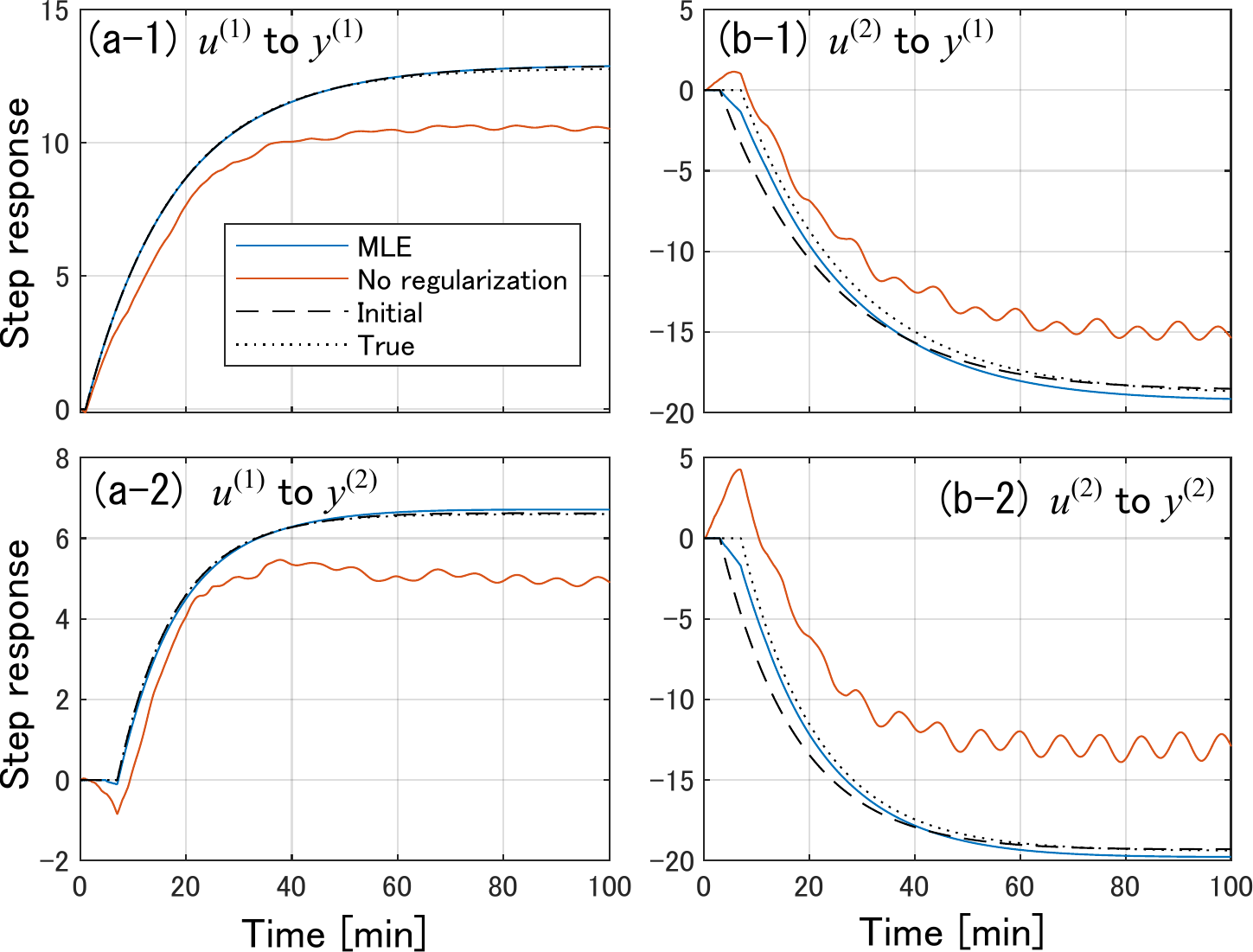}
  \caption{Step-response curves derived in experiment for ${\sf G}\sub{delay}$.
  Dashed line represents response of initial dynamics ${\sf G}_0$, dotted line true dynamics ${\sf G}\sub{delay}$, blue line ${\sf R}_{\lambda^*}$ estimated using MLE, and red line ${\sf R}_0$ estimated without regularization.}
  \label{fig:step_response_delay}	
  \end{center}
\end{figure}
Figure~\ref{fig:step_response_delay} shows step-response curves derived in the experiment for ${\sf G}\sub{delay}$, where the meaning of each line is as in \figref{fig:step_response_static}.
In all panels, MLE enabled ${\sf R}_{\lambda^*}$ to avoid undesired oscillation and static-gain mismatch, which were observed in the step response of ${\sf R}_0$ (red lines).
In panels (b-1) and (b-2), response curves of ${\sf R}_{\lambda^*}$ differ from those of ${\sf G}_0$ around $[3\,\mathrm{min},\ 7\,\mathrm{min}]$, indicating that MLE enables correction of transport delay.
To quantify the correction, we illustrate impulse-response curves in \figref{fig:Impulse_response_delay}, where ${\sf R}_{\lambda^*}$ is denoted with the blue line, ${\sf G}_0$ with the dashed line, and ${\sf G}\sub{delay}$ with the dotted line.
The peaks of the impulse-response curves shifted from the original one $t=3$\,min, which indicates the correction of transport delay.
Regarding transport delay as the peak time of the impulse-response curves, we calculated the transport delay at $7.6$\,min for both $y^{(1)}$ and $y^{(2)}$, close to the true delay of $7.0$\,min.

Through the above experiments, we demonstrated that an optimal regularization parameter can be found with 
MLE and showed the possibility of MLE to reduce static-gain mismatch and transport-delay mismatch,
which might only be corrected with regularization.
It should be noted that MLE can slightly induce undesired MPM;
see, e.g., the static-gain mismatch between $\hat{{\sf R}}_{\lambda^*}$ and ${\sf G}_0$ in \figref{fig:step_response_static}(b-2).
Recalling that $|\Delta K_{21}|$ was smaller than $|\Delta K_{11}|$,
regression coefficients in terms of $y^{(2)}$ were more regularized than $y^{(1)}$, which might cause the undesired mismatch. 
Thus, the accuracy of corrected models might be improved by introducing multiple regularization parameters $\lambda_1, \cdots, \lambda_p$
and replacing the penalty term in \eqref{eq:regression} with $\|\mathrm{diag}(\lambda_1, \cdots, \lambda_p) \Delta {\sf R}\|_{\mathcal{L}_1}$.

\section{Discussion}
\label{sec:discussion}
Here, we discuss the applicability of MLE.
Then we delineate its advantage and cost compared with the conventional re-identification frameworks that rely on test signals.

As implied above, MLE can be applied to closed-loop time-variant systems with the following assumptions:
\begin{enumerate}
  \item Reference signals intermittently change due to a routine operation.
  \item Timescale on which system parameters vary is greater than the interval over which reference signals change.
  \item Measurement data of process inputs and outputs are collected around the change in reference signal(s) for correcting MPM.
\end{enumerate}
From the second or third assumption, MLE does not suit real-time parameter estimation
\cite{aastrom2013adaptive},
celebrated for its application to flight control and ship steering.
On the other hand, all the assumptions can hold in both continuous and batch processes, such as batch reactors.
Thus, MLE is available to a range of industrial processes beyond the distillation column described in \secref{sec:experiment}.

The applicability of MLE is also related to the feasibility and convergence of MPC.
Under the significant change in dead time $L_i$,
MPC can fall infeasible without re-tuning.
In this case, it is necessary to estimate $L_i$ based on the corrected model
and to modify the prediction horizon $H\sub{p}$ and control horizon $H\sub{c}$ based on the estimated $L_i$.
Also, to improve the convergence of MPC,
it might be necessary to re-tune other parameters, such as weight matrices, based on the corrected model.
For industrial processes, this re-tuning is easily implemented by a body of tuning guidelines or defaults \cite{Forbes2015,alhajeri2020tuning,Schwenzer2021}.

The advantage of MLE compared with the conventional frameworks is in plant operation safety and MPC's maintainability.
In the conventional frameworks, the injection of test signals can violate operating limits of plants \cite{Forbes2015}.
Although test signals can be designed to ensure plant operation within its limits, this design involves the engineering services or exclusive closed-loop identification tools by MPC vendors,
leading to the high maintenance cost.
MLE does not require any experiment, thus avoiding the above issues.

The cost of MLE is that the frequency of MPM correction is limited to that of reference changing (per month).
In conventional frameworks,
if test signals are persistently injected into plants,
their models can be more frequently updated (per day or week)
so that their accuracy is kept higher than in the case of MLE.
However, since most MPC controllers are robustly designed in order to avoid severe damage due to MPM \cite{Forbes2015,Qin2003},
such a frequent update might be excessive for keeping MPC performance.

Considering both the advantage and cost, MLE has the potential to be a more economical solution than conventional frameworks.
We contend that MLE discovers a new research problem---{\em how frequently should a model be updated from economic aspects?}

\section{CONCLUSIONS AND FUTURE WORK}
\label{sec:conclusion}

We proposed a concept called MLE and its implementation, which enables us to mitigate the degradation in control performance in a non-invasive manner.
The purpose with MLE is to continually update process models as precisely as possible by using routine operating data when the timescale of the process aging is much larger than the interval of excitation of reference signals.
We implemented MLE via the $\mathcal{L}_1$ regularized regression and developed a methodology for finding its optimal parameter by using cross-validation.
Through numerical experiments for a pilot-scale distillation column controlled by an MPC controller, we showed that the optimal parameter for updating its model exists and can be found with MLE.
We then constructed the updated model for the parameter and showed the possibility to correct both static-gain mismatch and transport-delay mismatch without injecting excitation signals to process inputs.
Therefore, MLE can be used for pilot-scale distillation columns.

Future work lies in implementing MLE for a wide range of (possibly nonlinear) processes.
Technically, in this note we used $\mathcal{L}_1$ regularization to obtain sparse matrix ${\sf R}$ reflecting transport delays.
According to the magnitude of transport delays,
other sparse identification methods (e.g., elastic-net) might be chosen in MLE.
Also, as implied in \secref{sec:discussion}, it will be intriguing to investigate the long-term economic benefit of MLE compared with conventional re-identification frameworks that rely on the injection of test signals.

\section{ACKNOWLEDGMENTS}

The authors thank Mr. Kazunobu Morita, Dr. Yoichi Nonaka, and Mr. Yoshinori Mochizuki for their valuable discussion.
The first author would like to thank Dr. Morimasa Ogawa for suggesting the practice of MPC.


\bibliographystyle{IEEEtran}

\end{document}